# Stream Control Transmission Protocol Steganography

Wojciech Frączek, Wojciech Mazurczyk, Krzysztof Szczypiorski
Institute of Telecommunications
Warsaw University of Technology
Warsaw, Poland
e-mail: wfraczek@gmail.com,{wmazurczyk, ksz}@tele.pw.edu.pl

*Abstract*— Stream Control Transmission Protocol (SCTP) is a new transport layer protocol that is due to replace TCP (Transmission Control Protocol) and UDP (User Datagram Protocol) protocols in future IP networks. Currently, it is implemented in such operating systems like BSD, Linux, HP-UX or Sun Solaris. It is also supported in Cisco network devices operating system (Cisco IOS) and may be used in Windows. This paper describes potential steganographic methods that may be applied to SCTP and may pose a threat to network security. Proposed methods utilize new, characteristic SCTP features like multi-homing and multi-streaming. Identified new threats and suggested countermeasures may be used as a supplement to RFC 5062, which describes security attacks in SCTP protocol and can induce further standard modifications.

## I. INTRODUCTION

Steganographic techniques have been used for ages and dates back to the ancient Greece [4]. The aim of the steganographic communication back then and now, in modern applications, is the same: hide secret data (steganogram) in innocent looking cover and send it to the proper recipient which is aware of the information hiding procedure. In ideal situation the existence of hidden communication cannot be detected by third parties. What distinguishes historical steganographic methods from modern ones is, in fact, only the form of the cover (carrier) for secret data. Historical methods used human skin, wax tables or letters etc., nowadays rather digital media like pictures, audio, video which are transmitted using telecommunication networks were often used. Recent trend in steganography is utilization of the network protocols as a steganogram carrier by modifying content of the packets they use, time relations between these packets or hybrid solutions. All of the information hiding methods that may be used to exchange steganograms in telecommunication networks is described by the term network steganography which was originally introduced by Szczypiorski in 2003 [8]. Many steganographic methods have been proposed and analyzed, e.g. [1]-[4]. They should be treated as a threat to network security, because they may cause e.g. confidential information leakage. Steganography as a network threat was marginalized for few years but now not only security staff but even business and consulting firms are becoming continuously aware of the potential danger and possibilities it creates [10].

Knowledge of the information hiding procedure is helpful to develop countermeasures therefore, it is important to identify potential, previously unknown possibilities for covert communication. It is especially important when it comes to new network protocols that are forecasted to be widely deployed in future networks. For example, the detailed analysis of information hiding methods in IPv6 protocol header was presented by Lucena et al. [9]. The same case is with Stream Control Transmission Protocol (SCTP) [5] which is a transport layer protocol and its main role is similar to both popular protocols Transmission Control Protocol (TCP) and User Datagram Protocol (UDP). It provides some of the same service features of both, ensuring reliable, in-sequence transport of messages with congestion control. Nevertheless, there are certain advantages which make SCTP a candidate for a transport protocol in future IP networks – the main are that it is multi-streaming and multi-homing.

To authors' best knowledge, there are no steganographic methods proposed for SCTP protocol. However, information hiding methods that have been proposed for TCP and UDP protocols (e.g. utilizing free/unused or not strictly standard-defined fields) may be utilized as well due to several similarities between these transport layer protocols and SCTP. Steganographic methods for TCP and UDP protocols were described by Rowland [1] and Murdoch and Lewis [2] and very good surveys on hidden communication can be found in Zander et al. [3] and Petitcolas et al. [4].

The popularity of the SCTP is still growing as it has been already deployed in many important operating systems like BSD, Linux (the most popular is lksctp [13]), HP-UX or Sun Solaris and is supported Cisco network devices operating system (Cisco IOS) and even in Windows if the proper library is installed [11].

This paper can be treated as a supplement to RFC 5062 [12], which describes security attacks in SCTP protocol and current countermeasures. However, it does not include any information about steganography-based attacks and ways to prevent them. That is why, in this paper we identify new attack opportunities to network security for SCTP and propose detection and/or elimination techniques.

The rest of the paper is structured as follows. Section 2 gives brief overview of SCTP protocol. In Section 3 network steganography methods that are characteristic for SCTP protocol are presented. Section 4 provides possible detection and elimination solutions for proposed methods. Finally, Section 5 concludes our work.

## II. OVERVIEW OF SCTP PROTOCOL

SCTP [5] was defined by the IETF Signaling Transport (SIGTRAN) working group in 2000, and is maintained by the IETF Transport Area (TSVWG) working group. It was being developed for one specific reason - transportation of telephony signaling over IP-based networks. However, its features make it capable of being general purpose transport layer protocol ([5], [6]).

SCTP, like TCP, provides reliable, in-sequence data transport with congestion control, but it also eliminate limitations of TCP, which are more and more onerous in many applications. SCTP allows also to set order-of-arrival delivery of the data, which means that the data is delivered to the upper layer as soon as it is received (a sequence number is of no significance). Unordered transmission can be set for all messages or only for part of the messages depending on application need.

The SCTP Partial Reliability Extension, defined in [7], is a mechanism which allows to send not all data if it is not necessary, i.e. data, which were not correctly received but got out-of-date. Decision not to transmit some data is made by sender. He/she has to inform a receiver that some data will not be sent and receiver should treat this data like correctly received and acknowledged. Partial Reliability Extension and order-of-arrival delivery enable to use SCTP in many applications which are using UDP now.

In TCP all data is sent as a stream of bits with no boundaries between messages. This behavior requires that TCP-based applications have to do message framing and provide a buffer for incomplete messages from TCP agent. In SCTP, data is sent as separate messages passed by the upper layer. This feature makes SCTP-based applications easier to develop than TCP-based ones.

Each SCTP connection (which is called association in SCTP) can use one or more streams, which are unidirectional logical channels between SCTP endpoints. Order-of-transmission or order-of-arrival delivery of data is performed within each stream separately, not globally. If one of the streams is blocked (i.e. a packet is lost and receiver is waiting for it), it does not affect other streams. Benefit of using multiple streams is illustrated in Fig. 1.

User X sends four messages (A, B, C, D) to user Y. There are two requirements concerning delivery order of these messages. Message A must be delivered before message B, and message C must be delivered before message D. In TCP messages are sent in following order: A, B, C, D (1). If message A is lost (2), other messages, in spite of the correct reception, cannot be dispatched to the upper layer until message A is retransmitted and successfully received by user Y (3). In SCTP, using multi-streaming, messages can be divided into two streams. Messages A and B can be sent within stream 1, and messages C and D can be sent within stream 2 (4). If message A is lost (5), only message B cannot be passed to the upper layer until message A is received. Messages C and D can be delivered to the upper layer, since they are sent within different stream than messages A and B (6).

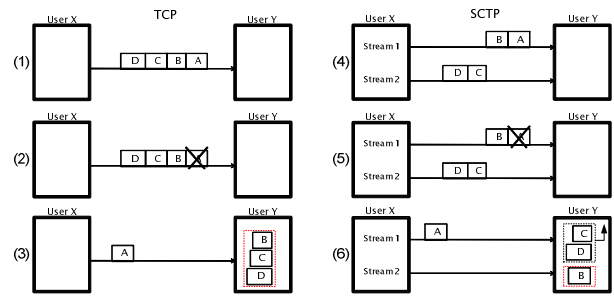

**Figure 1. Comparison of TCP and SCTP data transport using multiple streams**

Another SCTP feature is provision for protocol extensibility. Each SCTP packet consists of main header and one or more chunks (Fig. 2). There are two types of chunks: data chunks, which contain user data and control chunks, which are used to control data transfer. Each chunk consists of fields and parameters specific to chunk type (Fig. 3). Fields are mandatory, and parameters can be either mandatory or optional. SCTP packet structure allows defining not only new chunk types but also broadening functionality of the existing chunk types through defining new parameters.

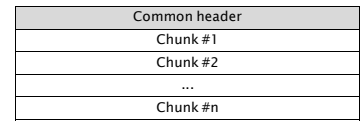

**Figure 2. SCTP packet format**

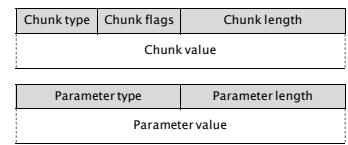

**Figure 3. SCTP chunks and parameters format**

SCTP supports multi-homing i.e. host ability to be visible in the network through more than one IP address, for instance if host is equipped with a few NICs (Network Interface Cards). Multi-homing in SCTP is used to provide more reliable data transfer. If there are no packets losses, all messages are transmitted using one source address and one destination address (primary path). If chunk is retransmitted, it should be sent using different path (different source and destination addresses) than primary path. Another advantage of SCTP multi-homing in SCTP is ability to failover data transfer if primary path is down.

SCTP uses a four-way handshake with cookie (Fig. 4), which provides protection against synchronization attack (type of Denial of Service attack) known from TCP. In SCTP, user initiates an association with INIT chunk. In response he/she receives INIT ACK chunk with cookie

(containing information that identifies proposed connection). Then he/she replies with a COOKIE ECHO with copy of received cookie. Reception of this chunk is acknowledged with COOKIE ACK chunk. After successful reception of COOKIE ACK association is established. Afterwards connected users can send data using DATA chunks and acknowledge reception of them with SACK chunks.

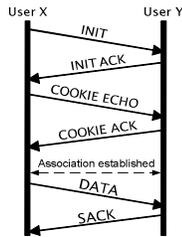

**Figure 4. SCTP association establishment**

Aside from described features, SCTP provide also built-in path MTU discovery, data fragmentation mechanism and, in general, it is considered more secure than TCP.

### III. SCTP-SPECIFIC STEGANOGRAPHIC METHODS AND DETECTION POSSIBILITIES

SCTP-specific steganographic methods can be divided in three groups:
- Methods that modify content of SCTP packets.
- Methods that modify how SCTP packets are exchanged.
- Methods that modify both content of SCTP and the way they are exchanged – hybrid methods.

#### A. Methods that modify content of SCTP packets

As mentioned before, each SCTP packet consists of chunks and each chunk can contain variable parameters. We propose 13 new steganographic methods which modify content of SCTP packets in the following chunks and parameters:
- INIT and INIT ACK chunks – used during initialization of SCTP association (methods I1, I2),
- DATA chunks – which contain user data (methods D1, D2),
- SACK chunks – used to acknowledge received DATA chunks (methods S1, S2),
- AUTH chunk – used to authenticate chunks (method A1),
- PAD chunk – used to pad packets (method P1),
- Variable parameters – used in specific chunks. (methods VP1-5).

Steganographic methods listed above are explained below.

INIT and INIT ACK chunks
(I1) *Initiate Tag* is a 32 bits value of *Verification Tag* field. It must be inserted into each SCTP packet, which is sent to the originator of INIT or INIT ACK chunks within this association. The *Initiate Tag* can be any value except 0, thus it may be used for steganographic purposes. Maximum bandwidth of this channel is 32 bits/chunk (fewer bits of this field should be used in order to limit chance of detection).

(I2) *Number of Inbound Streams* is a 16 bits field which define the maximum number of inbound streams that sender of the INIT or INIT ACK can handle within this association. In most cases, using more than one hundred streams is unlikely, thus at least a few the most significant bits can be used to insert hidden data. To limit the risk of detection not only the most significant bits may be used. Potential bandwidth of this method is 8 bits/chunk.

DATA chunks
(D1) *Stream Sequence Number* (SSN) is a 16 bits sequence number within each stream. If order-of-arrival delivery of data is set, there are no requirements concerning SSN. This feature makes it possible to use SSN to send steganograms. Maximum bandwidth of this channel is 16 bits/chunk. Presented method can be utilized only if unordered transmission is set for all data within a stream.

(D2) *Payload Protocol Identifier* is a 32 bits field which represents an upper layer protocol identifier. This field is not used by SCTP agent, it is for purposes of upper layer protocols. Value 0 indicates no identifier, other values should be standardized with IANA. SCTP does not verify this value, so it can be used to send secret data. Maximum bandwidth of this channel is 32 bits/chunk.

SACK chunks
(S1) *Advertised Receiver Window Credit* is a 32 bits field which indicates current size of the SACK sender's receiver buffer. A few least significant bits of this field can be utilized for steganographic purposes. Potential bandwidth of this method is 3-4 bits/chunk. It cannot be higher since it may affect flow control.

(S2) Duplicate TSNs, which are part of the SACK chunk, are sequence numbers of the duplicate chunks which has been received. This mechanism may enable hidden communication through adding not duplicating chunks TSNs to the list of duplicate TSNs. In spite of 32 bits length of TSN, potential steganographic bandwidth is few bits per chunk. This is because adding very different TSNs from recently sent is easy to detect. Presented method is harder to detect if it is used by multi-homed hosts since it should be considered to send SACK chunks with duplicates to other address than source address of DATA chunks.

AUTH chunks
(A1) *Shared Key Identifier* is a 16 bits field that indicates which pair of shared keys is used in this chunk. This field can be used for covert communication because receiver of the packet can authenticate sender through checking all previously exchanged shared keys. Potential steganographic bandwidth of this channel is 1-4 bits/chunk since, in most cases, there will be not many shared keys available. Detection of this method is quite hard because shared keys are established outside SCTP protocol.

PAD chunks

(P1) *Padding Data* is a field which length depends on padding needs. There are no requirements concerning value of this field, so it can be used for covert communication. Thus, steganographic bandwidth of this channel depends on size of padding data.

Variable Parameters

(VP1) IPv4 Address in *IPv4 Address Parameter* and IPv6 Address in *IPv6 Address Parameter* contain addresses of the sending endpoints. These parameters are used for multi-homed hosts and can be attached to INIT, INIT ACK and ASCONF (used to dynamic address reconfiguration) chunks. Each address in these parameters is considered as unconfirmed until its reachability is not checked. This behavior allows using these parameters for steganographic purposes by sending secret data instead of IP address. Maximum bandwidth is 32 bits/parameter for IPv4 address and 128 bits/parameter for IPv6 address.

(VP2) *Heartbeat Info Parameter* is used in HEARTBEAT chunk, which is exploited to verify reachability of the destination addresses. *Heartbeat Info Parameter* contains *Sender-Specific Heartbeat Info* field, which content is not defined, so it can be used a steganogram carrier. In Linux Kernel Stream Control Transmission Protocol (lksctp-2.6.28-1.0.10) implementation of SCTP, *Sender-Specific Heartbeat Info* field has 40 bytes, thus steganographic bandwidth for this methods is about 320 bits/chunk.

TABLE I. SUMMARY OF METHODS' POTENTIAL STEGANOGRAPHIC BANDWIDTH

| Steganographic method | Steganographic bandwidth | Units |
|---|---|---|
| I1 | 32 | bits/chunk |
| I2 | 8 | bits/chunk |
| D1 | 16 | bits/chunk |
| D2 | 32 | bits/chunk |
| S1 | 3-4 | bits/chunk |
| S2 | 3-4 | bits/chunk |
| A1 | 1-4 | bits/chunk |
| P1 | varies | n/a |
| VP1 | 32 | bits/par. |
| VP2 | 320 | bits/chunk |
| VP3 | 32 | bits/chunk |
| VP4 | 32 | bits/par. |
| VP5 | varies | n/a |

(VP3) *Random Number* in *Random Parameter* also can be used for covert communication. Steganographic bandwidth of this method depends on purpose of the number. If it is used in authentication process, random number has 32 bits and it is sent in INIT or INIT ACK chunks. That is why the maximum steganographic bandwidth is 32 bits/chunk.

(VP4) *ASCONF-Request Correlation ID* in *Add IP Address Parameter*, *Delete IP Address Parameter* and *Set Primary Address Parameter* is 32 bits field which identifies each request. The only requirement concerning its value is to be unique for each request, thus it may be used to transfer steganograms. The maximum steganographic bandwidth of this method is 32 bits/parameter.

(VP5) *Padding Data* in *Padding Parameter* can be exploited for covert communication in the same way as *Padding Data* in *Padding* chunk (see method P1). *Padding Parameter* can be used only in the INIT chunk.

## B. Methods that modify how SCTP packets are exchanged

MULTI-HOMING

SCTP multi-homing feature can be utilized to perform hidden communication. The main idea of the proposed steganographic method is presented in Fig. 4. Two users establish SCTP association (User 1 and User 2), each of them is equipped with more than one NIC. The primary path for the users' communication is through interfaces A and X (1). If $n_1$ denotes the number of the alternative sender NIC addresses (in Fig. 4 they are 2), and $n_2$ represents the number of alternative receiver NIC addresses (in Fig. 4 also 2) then each address can be used to represent one steganogram bit (or a sequence of bits). Possible alternative paths for communication between these users are: BY, BZ, CY and CZ. User 1's B interface IP address represents binary '0', C interface IP address binary '1' (similar situation is for User 2). Assigning the bits or sequence of bits to the users' NICs may depend on the IP addresses value i.e. available NICs addresses can be sorted from lowest to highest and then consecutive values (bit sequences) can be assigned to them.

If User 1 wants to send steganogram, he/she waits for the transmission error on primary path to occur and then retransmits chunk through appropriate path. For example, in Fig. 4, if User 1 wants to send steganogram which consists of the sequence '01', he/she waits for the transmission error on primary path to occur (1) and sends retransmitted packets through path BZ (2). Before sending steganogram it should be established which retransmitted chunks carry hidden data. Users can assume that all retransmissions carry bits of steganogram or should mark beginning of hidden communication, for example, with an initiation sequence (a sequence of retransmitted chunks through previously agreed paths).

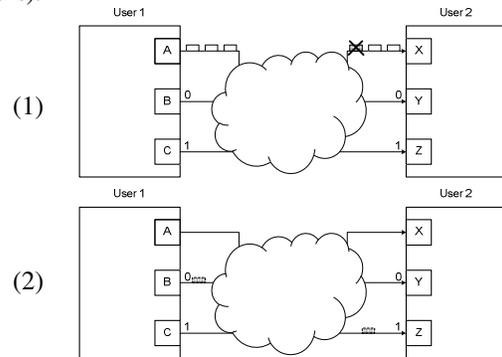

**Figure 4. Multi-homing based steganographic method**

Steganographic bandwidth $S_{B-MH}$ for this method can be expressed as

$$S_{B-MH} = \log_2(n_1) + \log_2(n_2) \quad [bits/chunk] \quad (3\text{-}1)$$

For example in Fig. 4, if SCTP packets rate is 250 packets/s, assuming that each packet contains only single data chunk and the retransmission rate is 2% (retransmission rate in Internet is up to 5%), then achieved steganographic bandwidth is 10 bits/s.

MULTI-STREAMING

In SCTP, multi-streaming (for ordered delivery) is realized by utilizing two identifiers: Stream Identifier (SI), to uniquely mark stream and Stream Sequence Number (SSN) to ensure correct order of packets at the receiver. Despite these two identifiers each DATA chunk contains also Transmission Sequence Number (TSN) that is assigned independently to each chunk.

Steganographic method that adopts multi-streaming is based on determined assignment of TSNs for every chunk distributed along different streams. SIs in subsequent DATA chunks will represent hidden data bits. The example for this method is presented in Fig. 5.

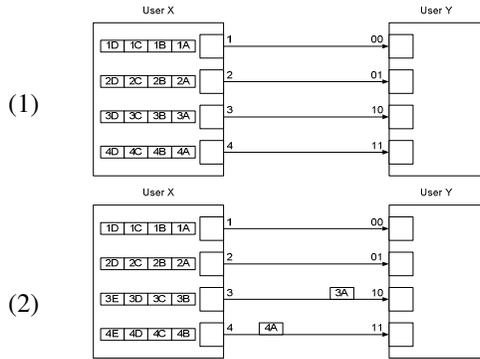

Figure 5. Multi-streaming based steganographic method

At initialization phase of the SCTP association users negotiate a number of utilized streams (in the example there are 4 streams). Each stream is assigned with binary sequences (1) – from '00' to '11'. Sending the data through certain stream depends on the steganogram bits. Therefore, if User X wants to secretly transfer '1011' bits sequence he/she first sends data through stream 3, then through stream 4 (2). If $s$ denotes the number of available streams, then maximum steganographic bandwidth $S_{B-MS}$ for this method may be expressed as

$$S_{B-MS} = \log_2(s) \quad [bits/chunk] \qquad (3\text{-}2)$$

For example, if we assume that the overt communication rate is 250 packets/s, each packet has only one chunk with data and 4 streams are used then the steganographic bandwidth is 500 bits/s.

C. Hybrid method

For SCTP partial reliability extension was also proposed by Stewart et al. [7]. It allows not retransmitting certain data despite the fact it was not successfully received. It is possible through the FORWARD TSN (FT) chunk, where new acknowledge TSN is inserted. After receiving such message receiving side treats missing chunks with equal or lower TSNs as they were properly delivered. This functionality may be adopted for steganographic purposes. The idea of the proposed method is similar in concept to LACK which was developed for real-time multimedia services by Mazurczyk and Szczypiorski [14].

The main idea of the proposed method is presented in Fig. 6.

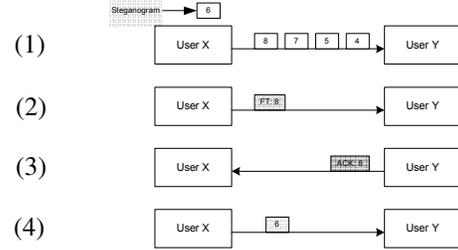

Figure 6. Multi-streaming based steganographic method

From the User X data sent chunk with TSN 6 is skipped and to this chunk steganogram is inserted (1). Next, User X sends FT chunk to signal new acknowledged TSN (2). After successful reception of FT chunk, User Y issues SACK chunk with new acknowledged TSN (3). When User X receives SACK chunk, he/she can send omitted DATA chunk with steganogram (4).

If we assume that the overt communication rate is 250 packets/s, each packet has only one chunk with payload size being 1000 bytes and we use 0.01% of packets to insert steganogram then the potential steganographic bandwidth is 200 bits/s.

IV. DETECTION POSSIBILITIES

For each of the groups of steganographic methods proposed in Section 3 detection or elimination solutions are sketched. The main aim of this Section is to point out potential enhancements that may be applied to SCTP standard to alleviate steganography utilization, ideally, at the standard development stage. Therefore, proposed countermeasures should be treated as guidelines for standard improvements.

A. Methods that modify content of SCTP packets

For steganographic methods that utilize modification to the SCTP packets content possible detection techniques and proposed countermeasures are depicted in Table II.

B. Methods that modify how SCTP packets are exchanged

MULTI-HOMING

It is worth noting that steganographic methods that utilize multi-homing are generally harder to detect than single-homing ones, because to detect covert communication it requires observing traffic on few, different communication paths.

Resistance to detection for method proposed in Section 3 depends on how future typical SCTP implementations will behave. If alternative paths for retransmitted chunks will often change proposed steganographic method that utilizes multi-homing will be harder to detect. But if retransmitted chunks will be send through only one alternative path then other behavior will be treated as anomaly. Thus, requirement that states that retransmitted chunks should be sent through only one alternative path should be enclosed in SCTP standard.

TABLE II. POSSIBLE STANDARD IMPROVEMENTS TO NEUTRALIZE STEGANOGRAPHIC METHODS THAT MODUFY CONTENT OF SCTP PACKETS

| Steg. method | Detection technique | Countermeasure (proposed standard change) |
|---|---|---|
| I1 | Analysis of Verification Tags values. | – |
| I2 | Comparison between values of Maximum Inbound Streams sent by "normal" users (users who do not use steganography) and suspicious user. | Limit possible values of Maximum Inbound Streams, i.e. only powers of 2 may be allowed. |
| D1 | Comparison between values of Stream Sequence Number sent by "normal" users (users who do not use steganography) and suspicious user. | For unordered transmission, Stream Sequence Number must be set to 0. |
| D2 | Checking value of Payload Stream Identifier. | Only standardized values must be allowed. |
| S1 | Analysis of a_rwnd values and sizes of received chunks. | – |
| S2 | Analysis of average number of duplicated chunks. | – |
| A1 | Analysis of Shared Key Identifier values. | Limit the number of shared keys for association to 1 or set one pair of shared keys for time slot, i.e. 10 minutes. |
| P1 | Analysis of Padding Data. | All bits of Padding Data must be set to 0. |
| VP1 | Checking the existence of IP addresses that are sent in these parameters. | Remove these parameters. Replace them with new chunk type, which will be sent from each user's address in order to add it to association. |
| VP2 | Comparison between values of Heartbeat Info Parameter sent by normal user (user who do not use steganography) and suspicious user. | Define value of Heartbeat Info Parameter. |
| VP3 | Analysis of Random Number. | – |
| VP4 | Comparison between values of ASCONF-Request Correlation ID sent by normal user (user who do not use steganography) and suspicious user. | ASCONF-Request Correlation ID must be a sequence number. |
| VP5 | Analysis of Padding Data. | All bits of Padding Data must be set to 0. |

Whatever the implementation, statistical analysis of NIC addresses used for retransmitted chunks may help to detect hidden communication.

Elimination of proposed steganographic method is possible by changing source and destination addresses of randomly chosen packet that contains retransmitted chunks.

MULTI-STREAMING

Similarly to the multi-homing based steganographic method detection of multi-streaming method may be hard to perform and depends on the concrete application were SCTP will be utilized. If the pattern of streams usage is established, then statistical SCTP traffic analysis may reveal hidden communication.

Elimination of the proposed steganographic method may be achieved by changing TSNs by an intermediate node e.g. edge router with steganography detection functionality. Such operation may successfully interrupt proper exchange of hidden data.

*C. Hybrid method*

If the number of intentionally omitted chunks is kept to the reasonable level then detection of such method is hard – statistical analysis of the frequency of moving acknowledged TSNs may be helpful.

Elimination of such method is possible by a specialized intermediate node which will be responsible for detection and dropping of chunks that have been already acknowledged by the receiver.

V. CONCLUSIONS

In this paper we presented sixteen different steganographic methods that can be used in SCTP protocol. All of these methods may lead to confidential information leakage and should be treated as a threat to network security. A lot of them may be evaded by changing SCTP standard – where it is possible certain improvements were proposed.

This analysis emphasizes how important it is to further inspect other network protocols that are to be utilized in future networks to avoid hidden communication as early as possible, ideally, still at the standard development stage.

REFERENCES

[1] C. Rowland, "Covert Channels in the TCP/IP Protocol Suite", First Monday, Peer Reviewed Journal on the Internet, July 1997

[2] Murdoch S.J., Lewis S., *Embedding Covert Channels into TCP/IP*, Information Hiding (2005), pp. 247-26

[3] S. Zander, G. Armitage, P. Branch, "A Survey of Covert Channels and Countermeasures in Computer Network Protocols", IEEE Communications Surveys & Tutorials, 3rd Quarter 2007, Volume: 9, Issue: 3, pp. 44-57, ISSN: 1553-877X

[4] Petitcolas F., Anderson R., Kuhn M., Information Hiding – A Survey: IEEE Special Issue on Protection of Multimedia Content, July 1999

[5] Stewart R.: Stream Control Transmission Protocol. RFC 4960, September 2007.

[6] Stewart R., Xie Q.: Stream Control Transmission Protocol (SCTP): A Reference Guide. Addison-Wesley, 2002.

[7] Stewart R., Ramalho M., Xie Q., Tuexen M., Conrad P.: Stream Control Transmission Protocol (SCTP) Partial Reliability Extension. RFC 3758, May 2004

[8] Szczypiorski K., Steganography in TCP/IP Networks. State of the Art and a Proposal of a New System – HICCUPS, Institute of Telecommunications' seminar, Warsaw University of Technology, Poland, November 2003

URL:http://krzysiek.tele.pw.edu.pl/pdf/steg-seminar-2003.pdf

[9] N. B. Lucena, G. Lewandowski, and S. J. Chapin, Covert Channels in IPv6, Proc. Privacy Enhancing Technologies (PET), May 2005, pp. 147–66.

[10] Frost & Sullivan, Steganography: Future of Information Hiding, Technical Insights Deliverable, December 2009

http://www.frost.com/prod/servlet/report-toc.pag?repid=D1D9-01-00-00-00

[11] SCTP library (sctplib): URL: http://www.sctp.de/sctp-download.html

[12] R. Stewart, M. Tuexen, G. Camarillo, Security Attacks Found Against the Stream Control Transmission Protocol (SCTP) and Current Countermeasures, RFC 5062, September 2007

[13] The Linux Kernel Stream Control Transmission Protocol (lksctp) project: http://lksctp.sourceforge.net/

[14] Mazurczyk W, Szczypiorski K (2008) Steganography of VoIP Streams, In: R. Meersman and Z. Tari (Eds.): OTM 2008, Part II - Lecture Notes in Computer Science (LNCS) 5332, Springer-Verlag Berlin Heidelberg, Proc. of The 3rd International Symposium on Information Security (IS'08), Monterrey, Mexico, November 2008, pp. 1001-1018